\documentclass[11pt,nofootinbib,floatfix,onecolumn,preprintnumbers]{revtex4}
\pdfoutput=1

\usepackage{amsmath,amssymb}
\usepackage{graphicx}
\usepackage{rotating}
\usepackage{color}
\usepackage{xcolor}
\usepackage{multirow}
\usepackage{fontenc}
\usepackage{slashed}
\usepackage{longtable}
\usepackage{dcolumn}
\usepackage{bm}
\usepackage{hyperref}
\hypersetup{colorlinks,bookmarksopen,bookmarksnumbered,linkcolor=blue,pdfstartview=FitH,urlcolor=blue,citecolor=blue}

\newcommand{\be}{\begin{equation}}
\newcommand{\ee}{\end{equation}}
\newcommand{\ba}{\begin{eqnarray}}
\newcommand{\ea}{\end{eqnarray}}
\newcommand{\bea}{\begin{eqnarray*}}
\newcommand{\eea}{\end{eqnarray*}}
\newcommand{\nn}{\nonumber}

\newcommand{\AddrKim}{%
Department of Physics and IPAP, Yonsei University, Seoul 120-749, Korea
}

\newcommand{\AddrGabriel}{%
Departamento de F\'isica, Centro de Investigaci\'on y de Estudios Avanzados del Instituto Polit\'ecnico Nacional,
Apartado Postal 14-740, 07000 M\'exico D.F., M\'exico
}

\newcommand{\AddrIFIC}{%
Instituto de F\'{\i}sica Corpuscular (CSIC-Universitat de Val\`{e}ncia),
Apdo. 22085, E-46071 Valencia, Spain
}

\preprint{IFIC/16-71}
\begin{document}

\title{Remarks on the Standard Model predictions for $R(D)$ and $R(D^*)$}

\author{C. S. Kim}\email{cskim@yonsei.ac.kr}
\affiliation{\AddrKim}

\author{G. Lopez-Castro}\email{glopez@fis.cinvestav.mx}
\affiliation{\AddrGabriel}

\author{S. L. Tostado}\email{stostado@fis.cinvestav.mx}
\affiliation{\AddrGabriel}

\author{A. Vicente}\email{avelino.vicente@ific.uv.es}
\affiliation{\AddrIFIC}

\begin{abstract}
Semileptonic $b \to c$ transitions, and in particular the ratios
$R(D^{(*)}) = \frac{\Gamma(B \rightarrow D^{(*)} \tau \nu )}{\Gamma(B
  \rightarrow D^{(*)} \ell \nu )}$, can be used to test the
universality of the weak interactions. In light of the recent
discrepancies between the experimental measurements of these
observables by BaBar, Belle and LHCb and the Standard Model predicted
values, we study the robustness of the latter. Our analysis reveals
that $R(D)$ might be enhanced by lepton mass effects associated to the
mostly unknown scalar form factor. In constrast, the Standard Model
prediction for $R(D^*)$ is found to be more robust, since possible
pollutions from $B^*$ contributions turn out to be negligibly small,
which indicates that $R(D^*)$ is a promising observable for searches
of new physics.
\end{abstract}

\maketitle

\section{Introduction}
\label{sec:intro}

Exclusive semileptonic $b\to c$ decays provide an ideal place to test
the quark flavor mixing structure
\cite{Cabibbo:1963yz,Kobayashi:1973fv} of the Standard Model (SM) and
to look for the existence of new charged currents. Among the required
theoretical ingredients, accurate calculations of the relevant
hadronic matrix elements are necessary to achieve these
goals. Observables like
\begin{equation}
R(D^{(*)}) = \frac{\Gamma(B \rightarrow D^{(*)} \tau \nu )}{\Gamma(B
  \rightarrow D^{(*)} \ell \nu )} \, ,
\end{equation}
with $\ell =e~\text{or}~\mu$, are particularly interesting to test the
universality of weak interactions, since most hadronic
  uncertainties cancel in these ratios.

The SM predictions we will be using for these ratios are shown in
Table \ref{table0}; the quoted (hadronic) uncertainties stem,
respectively, purely from lattice calculations \cite{HPQCD} and from
an estimate of higher order corrections to the ratio of $A_0/A_1$ form
factors in HQET \cite{fkn}. We should also mention that a more precise
SM prediction for $R(D)$ recently appeared in
\cite{Bigi:2016mdz}. Using unitarity constraints on the form factors
of $B \to D \ell \nu$ decays in three different models for the vector
and scalar form factors and fitting the resulting improved formula to
very recent experimental data and lattice results, the authors of this
reference found $R(D)=0.299(3)$, in good agreement with \cite{HPQCD}
and \cite{bkt}. The experimental situation regarding $R(D^{(*)})$ has
improved lately with new results from Belle \cite{Huschle:2015rga,
  Abdesselam:2016cgx} and LHCb \cite{Aaij:2015yra} collaborations, to
be added to previous results from BaBar \cite{Lees:2012xj}.  The
current world averages reported by the HFAG \cite{HFAG} exceed the SM
predictions by 1.9 $\sigma$ for $R(D)$ and 3.3 $\sigma$ for $R(D^*)$.

\begin{table}
\begin{tabular}{|c|c |c|}
\hline
Source & $R(D)$ & $R(D^*)$ \\
\hline
HFAG Exp. Av. \cite{HFAG} & \ $0.397\pm0.040\pm 0.028$ \ & \ $0.316 \pm 0.016\pm 0.010$\  \\
\hline
SM Prediction & $0.300 \pm 0.008$ \cite{HPQCD}& $0.252\pm0.003$ \cite{fkn}\\
\hline
\end{tabular}
\caption{\small Summary of experimental results and SM predictions for $R(D^{(*)})$.  }
\label{table0}
\end{table}

These hints of a possible violation of lepton universality have
prompted many theoretical proposals, which include the exchange of
charged scalars
\cite{Fajfer:2012jt,Crivellin:2012ye,Celis:2012dk,Bailey:2012jg,Tanaka:2012nw,Ko:2012sv,Freytsis:2015qca,Crivellin:2015hha},
leptoquarks (or, equivalently, R-parity violating supersymmetry)
\cite{Tanaka:2012nw,Deshpande:2012rr,Sakaki:2013bfa,Dorsner:2013tla,Alonso:2015sja,Bauer:2015knc,Barbieri:2015yvd,Fajfer:2015ycq,Freytsis:2015qca,Calibbi:2015kma,Hati:2015awg,Hati:2016thk,Deppisch:2016qqd,Zhu:2016xdg,Deshpand:2016cpw,Becirevic:2016yqi,Sahoo:2016pet,Hiller:2016kry,Bhattacharya:2016mcc,Das:2016vkr},
vector resonances \cite{Buttazzo:2016kid} or a $W^\prime$ boson
\cite{He:2012zp,Freytsis:2015qca,Greljo:2015mma,Boucenna:2016wpr,Boucenna:2016qad,Bhattacharya:2016mcc}. Possible
effects due to the presence of light sterile neutrinos have also been
explored in \cite{Abada:2013aba,Cvetic:2016fbv}. Proposals for
understanding the anomalies in the framework of an effective field
theory that incorporates dimension-6 scalar, vector and tensor
operators have also appeared in
\cite{Biancofiore:2013ki,Calibbi:2015kma,Bhattacharya:2015ida,Alonso:2016gym,Alok:2016qyh,Bardhan:2016uhr}.
We also note that the pQCD approach with lattice QCD input
\cite{Fan:2015kna} has shown drastically reduced discrepancies from
the experimental results.

In this letter we check the robustness of the SM prediction for the
$R(D^{(*)})$ ratios. While the vector form factor (VFF) predictions
for $B\to D\ell \nu$ decays have been tested with some detail in
measurements of the branching ratios and $q^2$-distributions for light
lepton channels, this is not the case for the scalar form factor (SFF)
which is visible only in decays with $\tau$ leptons. Small departures
of the SFF from lattice calculations can make compatible the SM
prediction with current measurements for $R(D)$. Inversely, we can
argue that the present experimental result for the $B\to D\tau \nu$
rate can determine the mostly unknown SFF, and test our knowledge of
nonperturbative QCD, instead.

In contrast, the compatibility of the SM with the observed value of
$R(D^*)$ would require unreasonably large departures from current form
factor calculations. Here we will show that if one defines $R(D^*)$
from a narrow window of the $D\pi$ invariant mass in $B\to D\pi\ell
\nu_{\ell}$, the additional $B^*$ pole contribution that pollutes this
decay gives a negligible small contribution. Although the individual
branching ratios are sensitive to the size of the chosen narrow
window, the ratio $R(D^*)$ turns out to be rather insensitive. Thus,
the robustness of the SM prediction for $R(D^*)$ indicates that this
observable is more promising for new physics searches in view of
current discrepancies.

\section{Semileptonic $B\to P$ transitions and $R(D)$}
\label{sec:BtoD}

The simplest semileptonic charged $B\to P$ transitions ($P=D,\pi$) are denoted by $B(p_B) \to P(p_P) \ell^-(p)\bar{\nu}_{\ell}(p')
\, $. The square of the momentum transfer $q=p_B-p_P=p+p'$ that characterizes the hadronic current varies within the interval
$m^2_{\ell} \leq q^2 \leq (m_B-m_P)^2$. Up to terms of $O(q^2/m_W^2)$, the tree-level decay amplitude is written as
\be
{\cal M}=\frac{G_F}{\sqrt{2}} V_{cb} \langle P| \bar{q}\gamma_{\mu}b| B \rangle \cdot \bar{u}(p)\gamma^{\mu}(1-\gamma_5) v(p')\ . \label{amplBP}
\ee

Lorentz covariance fixes the hadronic matrix element to have the form\footnote{The replacement $f_+(q^2)\to f_+(q^2)/(1-q^2/m_W^2)$ accounts for the finite mass of the $W$ gauge-boson. Its effects in $B\to D$ transition  are relevant only for precision studies aiming at an accuracy at the few per mille level.}
\be
\langle P| \bar{q}\gamma_{\mu}b| B \rangle =f_+(q^2) \left[ (p_B+p_P)_\mu-\frac{\Delta_{BP}}{q^2} q_{\mu}\right]+f_0(q^2) \frac{\Delta_{BP}}{q^2} q_{\mu} \,
\ee
where we have defined $\Delta_{BP}\equiv m_B^2-m_P^2$. The VFF and SFF are $f_+(q^2)$ and $f_0(q^2)$,
respectively. They are related at $q^2=0$ as
$f_+(0)=f_0(0)$.

The differential decay rate is given by
\be
\frac{d\Gamma(B\to P\ell \nu_{\ell})}{dq^2} = \frac{G_F^2|V_{cb}|^2}{192\pi^3 m_B^3} \eta_{\rm em} \left[c_{+}^{\ell}(q^2)\big |f_+(q^2)\big |^2+c_0^{\ell}(q^2)\big |f_0(q^2)\big|^2\right]\ ,
\ee
where $\eta_{\rm em}$ denotes the electroweak corrections \cite{HPQCD}.

\begin{figure}
\centering
\includegraphics[width=0.48\textwidth]{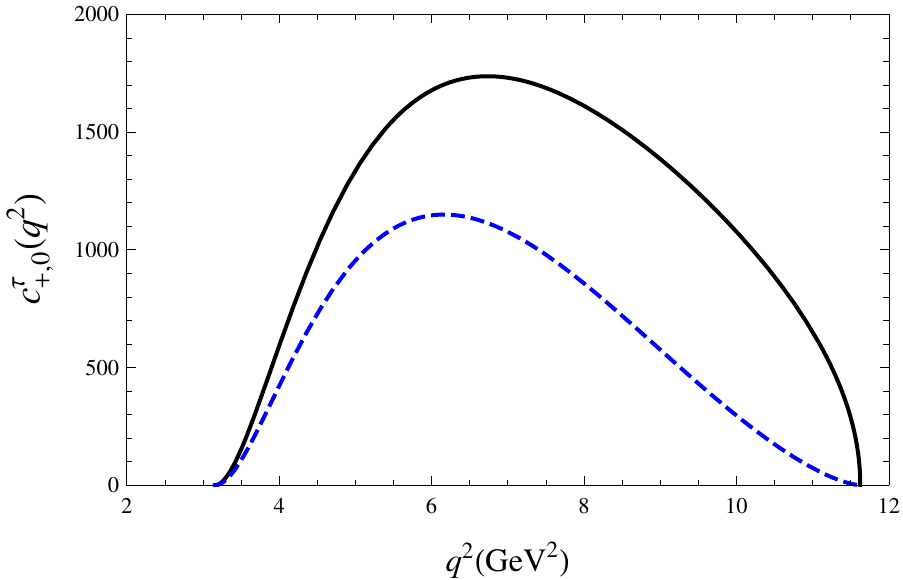}
\includegraphics[width=0.48\textwidth]{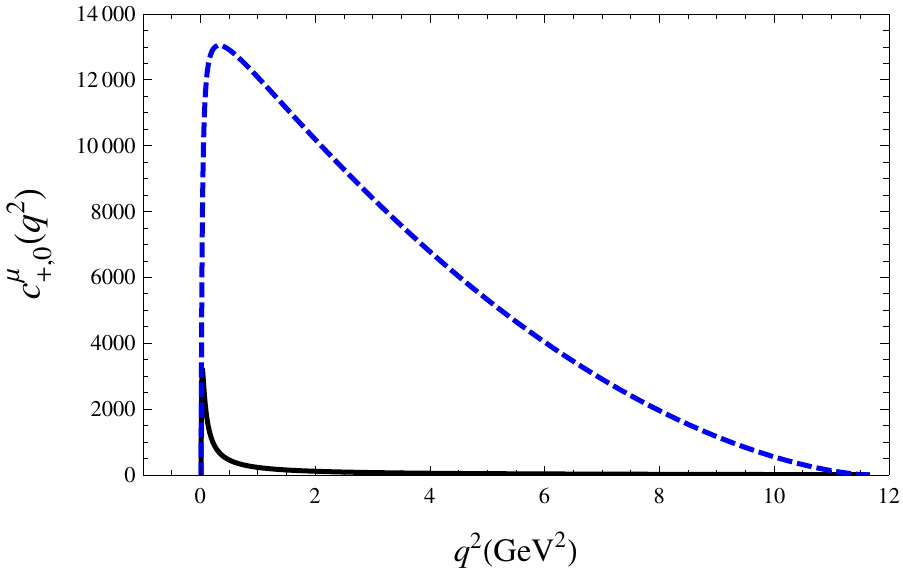}
\caption{\small  The $c_0^{\ell}(q^2)$ ($c_+^{\ell}(q^2)$) coefficient of $B\to D \ell^- \bar{\nu}_{\ell}$ decays are plotted as a function of $q^2$ with a solid (dashed) line. The left (right) panel corresponds to $\ell=\tau$ ($\ell=\mu$).}
\label{figu1}
\end{figure}

The coefficients $c_{+,0}^{\ell}$ that multiply the squared form factors in
the above expression are defined as
\begin{align}
c_+^\ell & = \lambda^{3/2} \left( q^2, m_B^2, m_P^2 \right) \left[ 1 - \frac{3}{2} \frac{m_\ell^2}{q^2} + \frac{1}{2} \left( \frac{m_\ell^2}{q^2} \right)^3 \right] \, , \\
c_0^\ell & = m_\ell^2 \lambda^{1/2} \left( q^2, m_B^2, m_P^2 \right) \frac{3}{2} \frac{m_B^4}{q^2} \left( 1 - \frac{m_\ell^2}{q^2} \right)^2 \left( 1 - \frac{m_P^2}{m_B^2} \right)^2 \, ,
\end{align}
where $\lambda \left( q^2, m_B^2, m_P^2 \right) = \left[ q^2 - (m_B +
  m_P)^2 \right]\left[ q^2 - (m_B - m_P)^2 \right]$, and are shown in
Figure \ref{figu1} for $\ell=\tau$ and $\ell=\mu$
\cite{bkt,Becirevic:2013zw}. These plots clearly show that the effects
of the SFF are sizable for the $B\to D\tau \nu$ transition, but
negligibly small for $B\to D \ell \nu$ decays; also, the effects of
the SFF are less important in the $B \to \pi \tau \nu$ transition.

\begin{figure}
\centering
\includegraphics[width=0.80\textwidth]{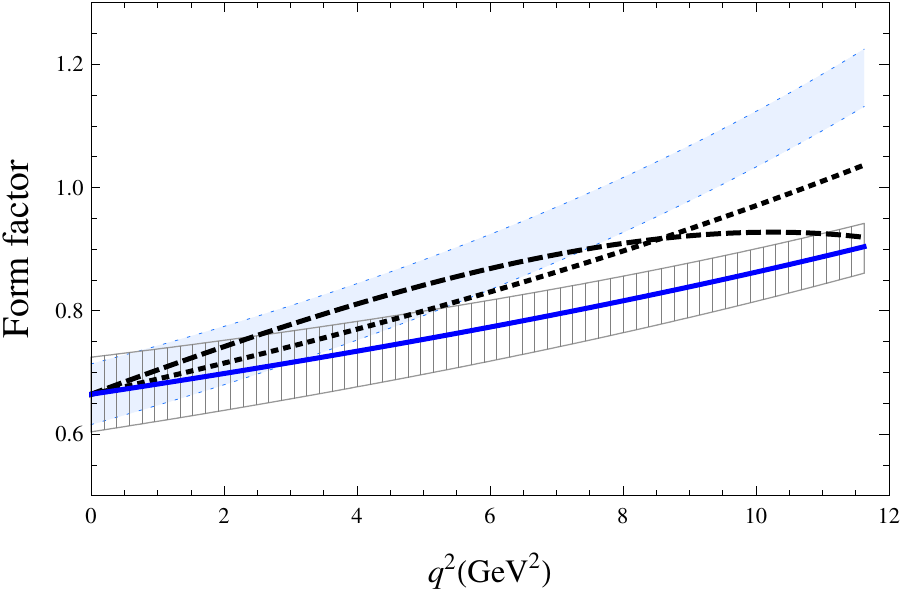}
\caption{\small The bands represent the Lattice predictions \cite{HPQCD} for the vector (light shaded) and scalar (vertical stripes)   form factors. The solid, dashed and dotted lines correspond to the scalar form factor obtained from the parametrization given in Eq. (\ref{quadratic}) with ($\alpha\  ($GeV$^{-2})$, $\beta\ ($GeV$^{-4})$)=$(-0.020,0.0),\ (+0.016,-0.003)$ and $(-0.008,-0.0002)$, respectively. }
\label{figu2}
\end{figure}

We focus now on the $P = D$ case. The vector and scalar form factors
calculated in Ref. \cite{HPQCD} using lattice QCD are shown in Figure
\ref{figu2}. The shaded bands represent the quoted errors in Ref.
\cite{HPQCD}. From the behavior of the form factors obtained from
lattice calculations and the condition $f_0(0)=f_+(0)$, the following
scaling relation \be f_0(q^2)=\left[1+\alpha \, q^2+\beta \, q^4
  \right]f_+(q^2) \, . \label{quadratic} \ee reproduces the scalar
form factor within the kinematical range of $B\to D$ transitions.  In
particular, the linear approximation $\alpha=-0.020(1)$ GeV$^{-2}$,
$\beta=0$ \cite{bkt,Becirevic:2013zw}, the solid line in Figure
\ref{figu2}, reproduces very well the central values for the SFF
obtained in lattice calculations \cite{HPQCD}.

By taking a different choice for the ($\alpha,\ \beta$) parameters one
can get a SM prediction closer to the measured value of the $R(D)$
observable. Since the lattice results are expected to be more reliable
at large $q^2$ values, one may choose $(\alpha,\beta)$ so that the
quadratic relation (\ref{quadratic}) and the lattice results for
$f_0(q^2)$ coincide at $q^2_{\rm max}$, as ilustrated by the dashed
line in Figure \ref{figu2}. Another possible choice is to allow the
scalar form factor to depart from its lattice QCD value at maximun
$q^2$, as shown by the dotted line in Figure \ref{figu2}.

These two possible departures from the linear scaling \cite{bkt,Becirevic:2013zw} between scalar and vector form factors
lead to similar values of $R(D)$, as shown in Table \ref{table1}, in better agreement with the experimental measurement.
We note that assuming an error bar for the dashed line as wide as the 
one for the lattice calculation of the SFF (vertical stripes band in 
Figure \ref{figu2}) would lead to an overlap among them. In this case, 
the resulting $R(D)$ value would be very close to the experimental 
measurement.

One may re-interpret this by stating that the current experimental
value of $R(D)$ indicates that the scalar form factor departs from
current lattice calculations by at least 10\% for
  certain $q^2$ values. This conclusion is justified by the absence of
  an independent and more precise test of the scalar form factor
  besides the one provided by measurements of $R(D)$. Measurements of
the $q^2$-distributions in $B\to D\tau\nu_{\tau}$ decays will then be
helpful as tests of lattice calculations. The robustness of the SM
calculation of $R(D)$ depends crucially on a better knowledge of the
SFF.

\begin{table}
\begin{tabular}{|c|c |c|}
\hline
& $\alpha$(GeV$^{-2}$), $\beta$(GeV$^{-4}$) & $R(D)$ \\
\hline\hline
-& $-0.020,\ 0.000$ & 0.300 \\
- &$-0.008, \ -0.0002$& 0.339 \\
-& $+0.016,\ -0.003$& 0.335 \\
\hline
SM Prediction \cite{pdg2014}& & 0.300 $\pm$ 0.008\\
Measured Value \cite{pdg2014}& & 0.397 $\pm$ 0.040 $\pm$ 0.028 \\
\hline
\end{tabular}
\caption{\small Predictions for the $R(D)$ ratio using the
  parametrization given in Eq. (\ref{quadratic}).}
\label{table1}
\end{table}

Finally, let us comment that the effect of similar changes in the SFF
are very small in the case of the ratio $R(\pi)=B(B\to
\pi\tau\nu_{\nu})/B(B\to \pi\mu\nu_{\mu})$. This follows from the
smaller difference between the $c_{0}^{\tau}$ and $c_{+}^{\tau}$
coefficients in $B\to \pi\tau\nu$ transitions.

\section{$B\to D\pi \ell^- \bar{\nu}_{\ell}$ decays and the definition of $R(D^*)$}
\label{sec:BtoDstar}

In contrast to the strong dependence of the $B \to D \tau \nu$ rate on the SFF, the $B \to D^* \tau \nu$ decay does not depend strongly on
any unknown form factors, and thus the SM prediction for $R(D^*)$ is rather robust. Therefore, a discrepancy between the $R(D^*)$ experimental measurement and
its predicted value in the SM would indicate a strong hint in favor of new physics. In the following we proceed to substantiate this claim. In particular, we will explore a possible deviation in $R(D^*)$ through an $R(D \pi)$ contribution, which we find to be negligibly small.

Theoretical calculations assume the $D^*$ meson in $B\to D^*\ell \nu_{\ell}$ (denoted as $B_{\ell 3}(D^*)$) decays to be an asymptotic state. This allows to assume that $\langle D^*|j_{\mu}|B\rangle$ is the hadronic matrix element of the $S$-matrix in the factorization approximation. Previous studies that take into account the effects of decays of $\tau$ leptons and/or $D^*$ mesons in some kinematical distributions were reported in \cite{Bordone:2016tex,Alonso:2016gym,Ligeti:2016npd}. Experimentally, the observable process is $B\to D\pi (D\gamma) \ell \nu$, and the observables associated to $B_{\ell 3}(D^*)$ are obtained by chosing $D\pi$ ($D\gamma$) events within a narrow window of their invariant mass distribution around the $D^*$ mass (hereafter we focus our discussion only on the $D\pi$ final states). The possible effects of higher $D^{**}$ resonances decaying into $D\pi$  are taken into account in simulations. Here we consider the possible effects of a $B^*$ pole contribution in addition to the $D^*$ pole and assess its effects in the extraction of the $R(D^*)$ observable.

\begin{figure}\centering
\includegraphics[width=14.0cm]{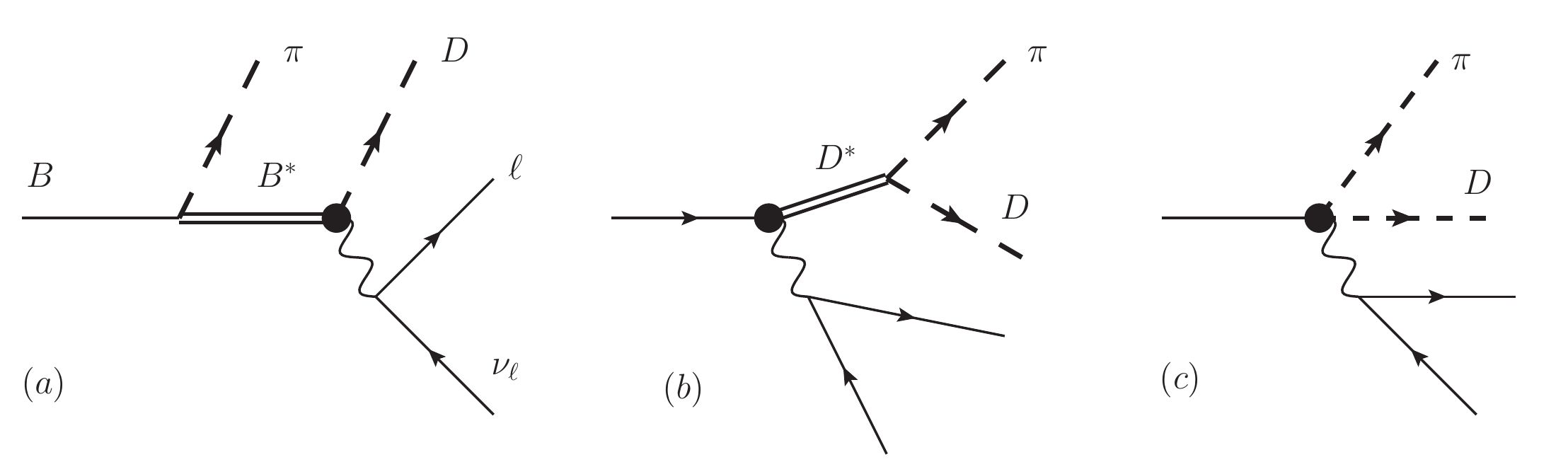}
  \caption{\small Contributions to the four-body semileptonic decay $B\to D\pi\ell^-\bar{\nu}_{\ell}$. Double-lines are used for the intermediate vector resonances.  The solid dot indicates the hadronic weak vertex. }\label{figu3}
\end{figure}

We can define the ratio
\be \label{rdpi}
R(D\pi) \equiv  \frac{\Gamma(B\to D\pi \tau \nu_{\tau})}{\Gamma(B\to D\pi \ell \nu_{\ell})}
\ee
from the $B (p_B)\to D(p_1) \pi(p_2) \ell(p_3) \nu(p_4)$ ($B_{\ell 4}(D\pi)$) decays. The Feynman diagrams contributing to this decay are shown in Figure \ref{figu3}.

In the narrow $D^*$ width approximation, the contribution in Figure
\ref{figu3}(b) yields $B(B\to D\pi\ell \nu_{\ell})=B(B\to D^* \ell
\nu_{\ell})\cdot B(D^* \to D\pi)$, thus the definition of $R(D\pi)$
and $R(D^*)$ are completely equivalent. This is not the case in the
presence of the other contributions, which will lead to
$R(D\pi)=R(D^*)\times(1+\delta_{D\pi})$, where $\delta_{D\pi}$ is a
pollution that remains owing to non-$D^*$ contributions. We expect
these additional contributions to be very small for a narrow window
around the $D^*$ mass, and we turn to evaluate them numerically. While
the $D^*$ pole gives rise to pure $p$-wave contributions of the $D\pi$
system, the $B^*$ pole can contribute to other configurations as well.

In our calculation we assume that the pole contributions are the dominant ones. Thus, the decay amplitude becomes:
\ba
{\cal M}&=& {\cal M}_a+ {\cal M}_b  \nn \\
&=& \frac{G_FV_{cb}}{\sqrt{2}} H_{\alpha}L^{\alpha}\ ,
\ea
where $L^{\alpha}$ is the leptonic weak current  and the hadronic matrix element is denoted as:
\be
H_{\alpha}= \langle D(p_1)\pi(p_2) | j_{\alpha}| B(p_B) \rangle \ ,
\ee
where $j_{\alpha}$ is the  SM weak current for the $b\to c$ transition.
Explicitly, the hadronic matrix elements corresponding to Figures \ref{figu3} (a) and (b) are:
\ba
H_{\alpha}^a&=&g_{BB^*\pi}\langle D(p_1) |j_{\alpha}|\widetilde{B^*}_{\beta}(p_B-p_2) \rangle (p_B+p_2)_{\nu} \left[ \frac{N^{\nu\beta}_{B^*}(p_B-p_2)}{(p_B-p_2)^2-m_{B^*}^2}\right] \\
H_{\alpha}^b&=& g_{DD^*\pi} \langle \widetilde{D^*}_{\beta}(p_1+p_2) |j_{\alpha}| B(p_B)\rangle (p_1-p_2)_{\nu} \left[ \frac{N^{\nu\beta}_{D^*}(p_1+p_2)}{(p_1+p_2)^2-m_{D^*}^2+im_{D^*}\Gamma_{D^*}}\right] ,
\ea
where $g_{VV^*\pi}$ denote the strong coupling constants and $N_{V^*}^{\nu\beta}(q) = T_{\nu\beta}(q)+L_{\nu\beta}(q)(m_{V^*}^2-q^2)/m_{V^*}^2$, with the transverse and longitudinal projectors defined by $T_{\nu\beta}(q)=g_{\nu\beta}-q_{\nu}q_{\beta}/q^2$ and $L_{\nu\beta}(q)=q_{\nu}q_{\beta}/q^2$. In the above expressions, the tildes denote the {\it off-shell} vector meson intermediate states with Lorentz index $\beta$ replacing their polarization four-vectors. The $D^*$ propagator has been provided with a finite width because it can be produced on-shell. Owing to similar $B^*-B$ and $D^*-D$ squared mass differences \cite{pdg2014}, the real parts in the denominators of the propagators may have similar sizes, thus the heavyness of the $B^*$ meson in principle does not provide a kinematical suppression.

For the purposes of numerical evaluations, we use the results of Ref. \cite{fkn} for the hadronic matrix element of the $B\to D^*$ transition. The $B^* \to D$ matrix element has a similar Lorentz structure as the $B\to D^*$ transition, although with different form factors; we can use the Heavy quark symmetry to relate it to the one of $B\to D$ decay, although for the purposes of the present work we use the form factors given in Ref. \cite{Chang:2016cdi}. We will use $g_{B B^*\pi}=20.0\pm 1.2$ \cite{Bernardoni:2014kla} which is consistent with other recent determinations \cite{Flynn:2015xna,ElBennich:2010ha}; also, we use the experimental value $g_{DD^*\pi}=8.39\pm 0.08$ \cite{pdg2014}.

 The four-body $B_{\ell 4}(D\pi)$ decay  can be described in terms of five independent kinematical variables. We chose the special set defined in Ref. \cite{FloresTlalpa:2005fz}, with $s_{12}=(p_1+p_2)^2$ and $s_{34}=(p_3+p_4)^2$ as two relevant variables. For our  example under consideration, the allowed phase space is determined by $(m_D+m_{\pi})^2 \leq s_{12}\leq (m_B-m_{\ell})^2$ and $m_{\ell}^2 \leq s_{34} \leq (m_B-\sqrt{s_{12}})^2$.  Leaving $s_{12}$ as the last integration variable, Eq. (\ref{rdpi}) can be written as:
\be
R_{D\pi}= \frac{  \int_{s_{12}^-}^{s_{12}^+} ds_{12} \frac{\textstyle d\Gamma (B\to D\pi\tau\nu)}{\textstyle ds_{12}}  } { \int_{s_{12}^-}^{s_{12}^+}  ds_{12}\frac{\textstyle d\Gamma (B\to D\pi\ell\nu)}{\textstyle ds_{12}} } \label{rdpicorrect}
\ee
where the limits of integration $s_{12}^{\pm}=(m_{D^*}\pm \Delta)^2$. This allows to study the dependence of the decay rates upon the size of the small window around the $D^*$ mass.

  As it was already mentioned, in the narrow width approximation obtained by setting $\Delta=0$, we recover the result $R(D\pi)=R(D^*)$ since the $B^*$ pole gives a vanishing contribution. In Table \ref{tableR}  we show the result of our calculations of the branching rations and of $R(D\pi)$ using Eq. (\ref{rdpicorrect}), for different values of $\Delta$. As it can be noticed, the branching fractions are sensitive to the cuts employed to define the $D^*$ mass window, although the ratio $R(D\pi)$ is insensitive to the value of $\Delta$ to the quoted accuracy. The relative size of the $B^*$ pole contribution with respect to the $D^*$pole contribution is very small for the different intervals chosen for $s_{12}$. Using the muon or the electron channels to normalize the $\tau$ decay rate, makes a difference of only about $0.5 \%$ in $R(D\pi)$.

\begin{table}\caption{\small Sensitivity of the branching ratio $B(\bar{B}^0 \to D^+\pi^0 \ell^- \nu)$ and $R_{\ell} (D\pi)$ (the subscript $\ell$ in the definition of $R(D\pi)$ refers to the specific  light $\ell$ channel used as normalization) to different regions of integration over $s_{12}$. In the calculation of the branching fractions we have used the $|V_{cb}|$ and form factor parameters of Ref. \cite{Dungel:2010uk} and the average $\bar{B}^0$ lifetime of Ref. \cite{pdg2014}. }
\centering
\begin{tabular}{|c|c|c|c|c|c|}
\hline
Channel\textbackslash$\Delta$ value & $0.5\Gamma_{D^*}$ &  $\Gamma_{D^*}$ & $1.5\Gamma_{D^*}$ & $ 2\Gamma_{D^*} $ &  1 MeV \\
\hline
$B(\bar{B}^0\to D^+\pi^0\tau^-\bar{\nu}_{\tau})$ & \ 0.00354 \ & \ 0.00499\ & \ 0.00563\ & \ 0.00598\  & \ 0.00689\    \\
\hline
$B(\bar{B}^0\to D^+\pi^0\mu^-\bar{\nu}_{\mu})$ & \ 0.01399\ & \ 0.01972\ & \ 0.02225 \ & \ 0.02360\ & \ 0.02725 \  \\
\hline
$B(\bar{B}^0\to D^+\pi^0e^-\bar{\nu}_{e})$ & \ 0.01405\ & \ 0.01981\ & \ 0.02235\   & \ 0.02372\ & \ 0.02736 \  \\
\hline
$R_{\mu}(D\pi) $&\ 0.2532  &\ 0.2532\ & \ 0.2532\ & \ 0.2534\ & \ 0.2532\  \\
$R_{e}(D\pi)$& \ 0.2520\  & \ 0.2520\ & \ 0.2520\ & \ 0.2520\ & \ 0.2520\  \\
\hline
\end{tabular}\label{tableR}
\end{table}

\section{Summary and conclusions}
\label{sec:sum}
The $R(D^{(*)})$ ratios are useful observables to study possible violations of the charged current universality. The SM prediction for $R(D)$ is sensitive to the scalar form factor  (SFF) in $B\to D$ semileptonic decays, and the current prediction relies on lattice calculations which have not been provided with independent tests. We have confirmed that increasing the SFF by up to  $10\%$ with respect to lattice results affects mainly the tau decay channels and can render the SM prediction in agreement with current measurements.

The situation is different for $R(D^*)$, since it requires strong variations of the SFF to produce a sizable change in the $B\to D^*\tau \nu_{\tau}$ rate. Since $D^*$ mesons are unstable states that are detected from $D\pi$ events very close to threshold in $B\to D\pi \ell \nu$ decays, we have studied the possible contamination of the $D^*$ signal by other allowed contributions. Considering the $B^*$ pole as the dominant additional contribution, we evaluate its impact in the extraction of $R(D^*)$ and find that it gives a negligible contribution when choosing a narrow window in the $D\pi$ invariant mass distribution. 

In conclusion, the SM prediction for $R(D^*)$ looks more robust than the one for $R(D)$ because it is less sensitive to hadronic form factors that are enhanced by lepton mass effects. Extracting the $R(D^*)$ ratio from observable $S$-matrix elements like $B\to D\pi \ell \nu_{\ell}$ may include additional $B^*$ contributions that pollute the $D^*$ signal; fortunately they turn out to be negligibly small.

\section*{Acknowledgements}

The work of CSK was supported in part by the NRF grant funded by Korea
government of the MEST (No. 2016R1D1A1A02936965). AV is grateful to
Damir Be\v{c}irevi\'c for fruitful discussions.  GLC is grateful to
A. Pich for useful conversations and thanks IFIC (Valencia) for its
hospitality during parts of this project. AV acknowledges financial
support from the ``Juan de la Cierva'' program (27-13-463B-731) funded
by the Spanish MINECO as well as from the Spanish grants
FPA2014-58183-P, Multidark CSD2009-00064, SEV-2014-0398 and
PROMETEOII/ 2014/084 (Generalitat Valenciana). GLC and SLT acknowledge
financial support from Conacyt (M\'exico) under projects 236394,
263916 and 296.

\bibliographystyle{utphys}
\bibliography{refs}

\providecommand{\href}[2]{#2}\begingroup\raggedright\begin{thebibliography}{10}

\bibitem{Cabibbo:1963yz}
N.~Cabibbo, ``{Unitary Symmetry and Leptonic Decays},''
\href{http://dx.doi.org/10.1103/PhysRevLett.10.531}{{\em Phys. Rev. Lett.}
  {\bfseries 10} (1963) 531--533}.

\bibitem{Kobayashi:1973fv}
M.~Kobayashi and T.~Maskawa, ``{CP Violation in the Renormalizable Theory of
  Weak Interaction},''
\href{http://dx.doi.org/10.1143/PTP.49.652}{{\em Prog. Theor. Phys.} {\bfseries
  49} (1973) 652--657}.

\bibitem{HPQCD}
{\bfseries HPQCD} Collaboration, H.~Na, C.~M. Bouchard, G.~P. Lepage,
  C.~Monahan, and J.~Shigemitsu, ``{$B \rightarrow D \ell \nu$ form factors at
  nonzero recoil and extraction of $|V_{cb}|$},''
  \href{http://dx.doi.org/10.1103/PhysRevD.93.119906,
  10.1103/PhysRevD.92.054510}{{\em Phys. Rev.} {\bfseries D92} no.~5, (2015)
  054510}, \href{http://arxiv.org/abs/1505.03925}{{\ttfamily arXiv:1505.03925
  [hep-lat]}}.
[Erratum: Phys. Rev.D93,no.11,119906(2016)].

\bibitem{fkn}
S.~Fajfer, J.~F. Kamenik, and I.~Ni\v{s}and\v{z}i\'c, ``{On the $B \to D^* \tau
  \bar \nu_{\tau}$ Sensitivity to New Physics},''
  \href{http://dx.doi.org/10.1103/PhysRevD.85.094025}{{\em Phys. Rev.}
  {\bfseries D85} (2012) 094025},
\href{http://arxiv.org/abs/1203.2654}{{\ttfamily arXiv:1203.2654 [hep-ph]}}.

\bibitem{Bigi:2016mdz}
D.~Bigi and P.~Gambino, ``{Revisiting $B\to D \ell \nu$},''
\href{http://arxiv.org/abs/1606.08030}{{\ttfamily arXiv:1606.08030 [hep-ph]}}.

\bibitem{bkt}
D.~Be\v{c}irevi\'c, N.~Ko\v{s}nik, and A.~Tayduganov, ``{$\bar B\to D\tau\bar
  \nu_\tau$ vs. $\bar B\to D\mu\bar \nu_\mu$},''
  \href{http://dx.doi.org/10.1016/j.physletb.2012.08.016}{{\em Phys. Lett.}
  {\bfseries B716} (2012) 208--213},
\href{http://arxiv.org/abs/1206.4977}{{\ttfamily arXiv:1206.4977 [hep-ph]}}.

\bibitem{Huschle:2015rga}
{\bfseries Belle} Collaboration, M.~Huschle {\em et~al.}, ``{Measurement of the
  branching ratio of $\bar{B} \to D^{(\ast)} \tau^- \bar{\nu}_\tau$ relative to
  $\bar{B} \to D^{(\ast)} \ell^- \bar{\nu}_\ell$ decays with hadronic tagging
  at Belle},'' \href{http://dx.doi.org/10.1103/PhysRevD.92.072014}{{\em Phys.
  Rev.} {\bfseries D92} no.~7, (2015) 072014},
\href{http://arxiv.org/abs/1507.03233}{{\ttfamily arXiv:1507.03233 [hep-ex]}}.

\bibitem{Abdesselam:2016cgx}
{\bfseries Belle} Collaboration, A.~Abdesselam {\em et~al.}, ``{Measurement of
  the branching ratio of $\bar{B}^0 \rightarrow D^{*+} \tau^- \bar{\nu}_{\tau}$
  relative to $\bar{B}^0 \rightarrow D^{*+} \ell^- \bar{\nu}_{\ell}$ decays
  with a semileptonic tagging method},''
\href{http://arxiv.org/abs/1603.06711}{{\ttfamily arXiv:1603.06711 [hep-ex]}}.

\bibitem{Aaij:2015yra}
{\bfseries LHCb} Collaboration, R.~Aaij {\em et~al.}, ``{Measurement of the
  ratio of branching fractions $\mathcal{B}(\bar{B}^0 \to
  D^{*+}\tau^{-}\bar{\nu}_{\tau})/\mathcal{B}(\bar{B}^0 \to
  D^{*+}\mu^{-}\bar{\nu}_{\mu})$},''
  \href{http://dx.doi.org/10.1103/PhysRevLett.115.159901,
  10.1103/PhysRevLett.115.111803}{{\em Phys. Rev. Lett.} {\bfseries 115}
  no.~11, (2015) 111803}, \href{http://arxiv.org/abs/1506.08614}{{\ttfamily
  arXiv:1506.08614 [hep-ex]}}.
[Addendum: Phys. Rev. Lett.115,no.15,159901(2015)].

\bibitem{Lees:2012xj}
{\bfseries BaBar} Collaboration, J.~P. Lees {\em et~al.}, ``{Evidence for an
  excess of $\bar{B} \to D^{(*)} \tau^-\bar{\nu}_\tau$ decays},''
  \href{http://dx.doi.org/10.1103/PhysRevLett.109.101802}{{\em Phys. Rev.
  Lett.} {\bfseries 109} (2012) 101802},
\href{http://arxiv.org/abs/1205.5442}{{\ttfamily arXiv:1205.5442 [hep-ex]}}.

\bibitem{HFAG}
``Heavy flavor averaging group.''
  \url{http://www.slac.stanford.edu/xorg/hfag/semi/winter16/winter16_dtaunu.html}.

\bibitem{Fajfer:2012jt}
S.~Fajfer, J.~F. Kamenik, I.~Ni\v{s}and\v{z}i\'c, and J.~Zupan, ``{Implications
  of Lepton Flavor Universality Violations in B Decays},''
  \href{http://dx.doi.org/10.1103/PhysRevLett.109.161801}{{\em Phys. Rev.
  Lett.} {\bfseries 109} (2012) 161801},
\href{http://arxiv.org/abs/1206.1872}{{\ttfamily arXiv:1206.1872 [hep-ph]}}.

\bibitem{Crivellin:2012ye}
A.~Crivellin, C.~Greub, and A.~Kokulu, ``{Explaining $B\to D\tau\nu$, $B\to
  D^*\tau\nu$ and $B\to \tau\nu$ in a 2HDM of type III},''
  \href{http://dx.doi.org/10.1103/PhysRevD.86.054014}{{\em Phys. Rev.}
  {\bfseries D86} (2012) 054014},
\href{http://arxiv.org/abs/1206.2634}{{\ttfamily arXiv:1206.2634 [hep-ph]}}.

\bibitem{Celis:2012dk}
A.~Celis, M.~Jung, X.-Q. Li, and A.~Pich, ``{Sensitivity to charged scalars in
  $\boldsymbol{B\to D^{(*)}\tau\nu_\tau}$ and $\boldsymbol{B\to\tau\nu_\tau}$
  decays},'' \href{http://dx.doi.org/10.1007/JHEP01(2013)054}{{\em JHEP}
  {\bfseries 01} (2013) 054},
\href{http://arxiv.org/abs/1210.8443}{{\ttfamily arXiv:1210.8443 [hep-ph]}}.

\bibitem{Bailey:2012jg}
J.~A. Bailey {\em et~al.}, ``{Refining new-physics searches in $B \to D \tau
  \nu$ decay with lattice QCD},''
  \href{http://dx.doi.org/10.1103/PhysRevLett.109.071802}{{\em Phys. Rev.
  Lett.} {\bfseries 109} (2012) 071802},
\href{http://arxiv.org/abs/1206.4992}{{\ttfamily arXiv:1206.4992 [hep-ph]}}.

\bibitem{Tanaka:2012nw}
M.~Tanaka and R.~Watanabe, ``{New physics in the weak interaction of $\bar B\to
  D^{(*)}\tau\bar\nu$},''
  \href{http://dx.doi.org/10.1103/PhysRevD.87.034028}{{\em Phys. Rev.}
  {\bfseries D87} no.~3, (2013) 034028},
\href{http://arxiv.org/abs/1212.1878}{{\ttfamily arXiv:1212.1878 [hep-ph]}}.

\bibitem{Ko:2012sv}
P.~Ko, Y.~Omura, and C.~Yu, ``{$B \to D^{(*)} \tau \nu$ and $B \to \tau \nu$ in
  chiral U(1)' models with flavored multi Higgs doublets},''
  \href{http://dx.doi.org/10.1007/JHEP03(2013)151}{{\em JHEP} {\bfseries 03}
  (2013) 151},
\href{http://arxiv.org/abs/1212.4607}{{\ttfamily arXiv:1212.4607 [hep-ph]}}.

\bibitem{Freytsis:2015qca}
M.~Freytsis, Z.~Ligeti, and J.~T. Ruderman, ``{Flavor models for $\bar{B} \to
  D^{(*)} \tau \bar{\nu}$},''
  \href{http://dx.doi.org/10.1103/PhysRevD.92.054018}{{\em Phys. Rev.}
  {\bfseries D92} no.~5, (2015) 054018},
\href{http://arxiv.org/abs/1506.08896}{{\ttfamily arXiv:1506.08896 [hep-ph]}}.

\bibitem{Crivellin:2015hha}
A.~Crivellin, J.~Heeck, and P.~Stoffer, ``{A perturbed lepton-specific
  two-Higgs-doublet model facing experimental hints for physics beyond the
  Standard Model},''
  \href{http://dx.doi.org/10.1103/PhysRevLett.116.081801}{{\em Phys. Rev.
  Lett.} {\bfseries 116} no.~8, (2016) 081801},
\href{http://arxiv.org/abs/1507.07567}{{\ttfamily arXiv:1507.07567 [hep-ph]}}.

\bibitem{Deshpande:2012rr}
N.~G. Deshpande and A.~Menon, ``{Hints of R-parity violation in B decays into
  $\tau \nu$},'' \href{http://dx.doi.org/10.1007/JHEP01(2013)025}{{\em JHEP}
  {\bfseries 01} (2013) 025},
\href{http://arxiv.org/abs/1208.4134}{{\ttfamily arXiv:1208.4134 [hep-ph]}}.

\bibitem{Sakaki:2013bfa}
Y.~Sakaki, M.~Tanaka, A.~Tayduganov, and R.~Watanabe, ``{Testing leptoquark
  models in $\bar B \to D^{(*)} \tau \bar\nu$},''
  \href{http://dx.doi.org/10.1103/PhysRevD.88.094012}{{\em Phys. Rev.}
  {\bfseries D88} no.~9, (2013) 094012},
\href{http://arxiv.org/abs/1309.0301}{{\ttfamily arXiv:1309.0301 [hep-ph]}}.

\bibitem{Dorsner:2013tla}
I.~Dor\v{s}ner, S.~Fajfer, N.~Ko\v{s}nik, and I.~Ni\v{s}and\v{z}i\'c,
  ``{Minimally flavored colored scalar in $\bar B \to D^{(*)} \tau \bar \nu$
  and the mass matrices constraints},''
  \href{http://dx.doi.org/10.1007/JHEP11(2013)084}{{\em JHEP} {\bfseries 11}
  (2013) 084},
\href{http://arxiv.org/abs/1306.6493}{{\ttfamily arXiv:1306.6493 [hep-ph]}}.

\bibitem{Alonso:2015sja}
R.~Alonso, B.~Grinstein, and J.~Martin~Camalich, ``{Lepton universality
  violation and lepton flavor conservation in $B$-meson decays},''
  \href{http://dx.doi.org/10.1007/JHEP10(2015)184}{{\em JHEP} {\bfseries 10}
  (2015) 184},
\href{http://arxiv.org/abs/1505.05164}{{\ttfamily arXiv:1505.05164 [hep-ph]}}.

\bibitem{Bauer:2015knc}
M.~Bauer and M.~Neubert, ``{Minimal Leptoquark Explanation for the
  R$_{D^{(*)}}$ , R$_K$ , and $(g-2)_g$ Anomalies},''
  \href{http://dx.doi.org/10.1103/PhysRevLett.116.141802}{{\em Phys. Rev.
  Lett.} {\bfseries 116} no.~14, (2016) 141802},
\href{http://arxiv.org/abs/1511.01900}{{\ttfamily arXiv:1511.01900 [hep-ph]}}.

\bibitem{Barbieri:2015yvd}
R.~Barbieri, G.~Isidori, A.~Pattori, and F.~Senia, ``{Anomalies in $B$-decays
  and $U(2)$ flavour symmetry},''
  \href{http://dx.doi.org/10.1140/epjc/s10052-016-3905-3}{{\em Eur. Phys. J.}
  {\bfseries C76} no.~2, (2016) 67},
\href{http://arxiv.org/abs/1512.01560}{{\ttfamily arXiv:1512.01560 [hep-ph]}}.

\bibitem{Fajfer:2015ycq}
S.~Fajfer and N.~Ko\v{s}nik, ``{Vector leptoquark resolution of $R_K$ and
  $R_{D^{(*)}}$ puzzles},''
  \href{http://dx.doi.org/10.1016/j.physletb.2016.02.018}{{\em Phys. Lett.}
  {\bfseries B755} (2016) 270--274},
\href{http://arxiv.org/abs/1511.06024}{{\ttfamily arXiv:1511.06024 [hep-ph]}}.

\bibitem{Calibbi:2015kma}
L.~Calibbi, A.~Crivellin, and T.~Ota, ``{Effective Field Theory Approach to $b
  \to s \ell \ell^{(\prime)}$, $B \to K^{(*)} \nu \bar \nu$ and $B \to D^{(*)}
  \tau \nu$ with Third Generation Couplings},''
  \href{http://dx.doi.org/10.1103/PhysRevLett.115.181801}{{\em Phys. Rev.
  Lett.} {\bfseries 115} (2015) 181801},
\href{http://arxiv.org/abs/1506.02661}{{\ttfamily arXiv:1506.02661 [hep-ph]}}.

\bibitem{Hati:2015awg}
C.~Hati, G.~Kumar, and N.~Mahajan, ``{$\bar{B}\rightarrow D^{(\ast)}\tau
  \bar{\nu}$ excesses in ALRSM constrained from $B$, $D$ decays and
  $D^{0}-\bar{D}^{0}$ mixing},''
  \href{http://dx.doi.org/10.1007/JHEP01(2016)117}{{\em JHEP} {\bfseries 01}
  (2016) 117},
\href{http://arxiv.org/abs/1511.03290}{{\ttfamily arXiv:1511.03290 [hep-ph]}}.

\bibitem{Hati:2016thk}
C.~Hati, ``{Explaining the diphoton excess in Alternative Left-Right Symmetric
  Model},'' \href{http://dx.doi.org/10.1103/PhysRevD.93.075002}{{\em Phys.
  Rev.} {\bfseries D93} no.~7, (2016) 075002},
\href{http://arxiv.org/abs/1601.02457}{{\ttfamily arXiv:1601.02457 [hep-ph]}}.

\bibitem{Deppisch:2016qqd}
F.~F. Deppisch, S.~Kulkarni, H.~P{\"a}s, and E.~Schumacher, ``{Leptoquark
  patterns unifying neutrino masses, flavor anomalies and the diphoton
  excess},''
\href{http://arxiv.org/abs/1603.07672}{{\ttfamily arXiv:1603.07672 [hep-ph]}}.

\bibitem{Zhu:2016xdg}
J.~Zhu, H.-M. Gan, R.-M. Wang, Y.-Y. Fan, Q.~Chang, and Y.-G. Xu, ``{Probing
  the R-parity violating supersymmetric effects in the exclusive $b\to
  c\ell^-\bar{\nu}_\ell$ decays},''
  \href{http://dx.doi.org/10.1103/PhysRevD.93.094023}{{\em Phys. Rev.}
  {\bfseries D93} no.~9, (2016) 094023},
\href{http://arxiv.org/abs/1602.06491}{{\ttfamily arXiv:1602.06491 [hep-ph]}}.

\bibitem{Deshpand:2016cpw}
N.~G. Deshpande and X.-G. He, ``{Consequences of R-Parity violating
  interactions for anomalies in $\bar B\to D^{(*)} \tau \bar \nu$ and $b\to s
  \mu^+\mu^-$},''
\href{http://arxiv.org/abs/1608.04817}{{\ttfamily arXiv:1608.04817 [hep-ph]}}.

\bibitem{Becirevic:2016yqi}
D.~Be\v{c}irevi\'c, S.~Fajfer, N.~Ko\v{s}nik, and O.~Sumensari, ``{Leptoquark
  model to explain the $B$-physics anomalies, $R_K$ and $R_D$},''
\href{http://arxiv.org/abs/1608.08501}{{\ttfamily arXiv:1608.08501 [hep-ph]}}.

\bibitem{Sahoo:2016pet}
S.~Sahoo, R.~Mohanta, and A.~K. Giri, ``{Explaining $R_{K}$ and $R_{D^{(*)}}$
  anomalies with vector leptoquark},''
\href{http://arxiv.org/abs/1609.04367}{{\ttfamily arXiv:1609.04367 [hep-ph]}}.

\bibitem{Hiller:2016kry}
G.~Hiller, D.~Loose, and K.~Sch{\"o}nwald, ``{Leptoquark Flavor Patterns \& B
  Decay Anomalies},''
\href{http://arxiv.org/abs/1609.08895}{{\ttfamily arXiv:1609.08895 [hep-ph]}}.

\bibitem{Bhattacharya:2016mcc}
B.~Bhattacharya, A.~Datta, J.-P. Gu\'evin, D.~London, and R.~Watanabe,
  ``{Simultaneous Explanation of the $R_K$ and $R_{D^{(*)}}$ Puzzles: a Model
  Analysis},''
\href{http://arxiv.org/abs/1609.09078}{{\ttfamily arXiv:1609.09078 [hep-ph]}}.

\bibitem{Das:2016vkr}
D.~Das, C.~Hati, G.~Kumar, and N.~Mahajan, ``{Towards a unified explanation of
  $R_{D^{(\ast)}}$, $R_{K}$ and $(g-2)_{\mu}$ anomalies in a left-right model
  with leptoquarks},'' \href{http://dx.doi.org/10.1103/PhysRevD.94.055034}{{\em
  Phys. Rev.} {\bfseries D94} (2016) 055034},
\href{http://arxiv.org/abs/1605.06313}{{\ttfamily arXiv:1605.06313 [hep-ph]}}.

\bibitem{Buttazzo:2016kid}
D.~Buttazzo, A.~Greljo, G.~Isidori, and D.~Marzocca, ``{Toward a coherent
  solution of diphoton and flavor anomalies},''
  \href{http://dx.doi.org/10.1007/JHEP08(2016)035}{{\em JHEP} {\bfseries 08}
  (2016) 035},
\href{http://arxiv.org/abs/1604.03940}{{\ttfamily arXiv:1604.03940 [hep-ph]}}.

\bibitem{He:2012zp}
X.-G. He and G.~Valencia, ``{$B$ decays with $\tau$ leptons in nonuniversal
  left-right models},''
  \href{http://dx.doi.org/10.1103/PhysRevD.87.014014}{{\em Phys. Rev.}
  {\bfseries D87} no.~1, (2013) 014014},
\href{http://arxiv.org/abs/1211.0348}{{\ttfamily arXiv:1211.0348 [hep-ph]}}.

\bibitem{Greljo:2015mma}
A.~Greljo, G.~Isidori, and D.~Marzocca, ``{On the breaking of Lepton Flavor
  Universality in B decays},''
  \href{http://dx.doi.org/10.1007/JHEP07(2015)142}{{\em JHEP} {\bfseries 07}
  (2015) 142},
\href{http://arxiv.org/abs/1506.01705}{{\ttfamily arXiv:1506.01705 [hep-ph]}}.

\bibitem{Boucenna:2016wpr}
S.~M. Boucenna, A.~Celis, J.~Fuentes-Mart\'in, A.~Vicente, and J.~Virto,
  ``{Non-abelian gauge extensions for B-decay anomalies},''
  \href{http://dx.doi.org/10.1016/j.physletb.2016.06.067}{{\em Phys. Lett.}
  {\bfseries B760} (2016) 214--219},
\href{http://arxiv.org/abs/1604.03088}{{\ttfamily arXiv:1604.03088 [hep-ph]}}.

\bibitem{Boucenna:2016qad}
S.~M. Boucenna, A.~Celis, J.~Fuentes-Mart\'in, A.~Vicente, and J.~Virto,
  ``{Phenomenology of an $SU(2) \times SU(2) \times U(1)$ model with
  lepton-flavour non-universality},''
\href{http://arxiv.org/abs/1608.01349}{{\ttfamily arXiv:1608.01349 [hep-ph]}}.

\bibitem{Abada:2013aba}
A.~Abada, A.~M. Teixeira, A.~Vicente, and C.~Weiland, ``{Sterile neutrinos in
  leptonic and semileptonic decays},''
  \href{http://dx.doi.org/10.1007/JHEP02(2014)091}{{\em JHEP} {\bfseries 02}
  (2014) 091},
\href{http://arxiv.org/abs/1311.2830}{{\ttfamily arXiv:1311.2830 [hep-ph]}}.

\bibitem{Cvetic:2016fbv}
G.~Cveti\v{c} and C.~S. Kim, ``{Rare decays of B mesons via on-shell sterile
  neutrinos},'' \href{http://dx.doi.org/10.1103/PhysRevD.94.053001}{{\em Phys.
  Rev.} {\bfseries D94} no.~5, (2016) 053001},
\href{http://arxiv.org/abs/1606.04140}{{\ttfamily arXiv:1606.04140 [hep-ph]}}.

\bibitem{Biancofiore:2013ki}
P.~Biancofiore, P.~Colangelo, and F.~De~Fazio, ``{On the anomalous enhancement
  observed in $B \to D^{(*)}\tau{\bar \nu}_\tau$ decays},''
  \href{http://dx.doi.org/10.1103/PhysRevD.87.074010}{{\em Phys. Rev.}
  {\bfseries D87} no.~7, (2013) 074010},
\href{http://arxiv.org/abs/1302.1042}{{\ttfamily arXiv:1302.1042 [hep-ph]}}.

\bibitem{Bhattacharya:2015ida}
S.~Bhattacharya, S.~Nandi, and S.~K. Patra, ``{Optimal-observable analysis of
  possible new physics in $B\to D^{(\ast)}\tau\nu_{\tau}$},''
  \href{http://dx.doi.org/10.1103/PhysRevD.93.034011}{{\em Phys. Rev.}
  {\bfseries D93} no.~3, (2016) 034011},
\href{http://arxiv.org/abs/1509.07259}{{\ttfamily arXiv:1509.07259 [hep-ph]}}.

\bibitem{Alonso:2016gym}
R.~Alonso, A.~Kobach, and J.~Martin~Camalich, ``{New physics in the kinematic
  distributions of $\bar B\to
  D^{(*)}\tau^-(\to\ell^-\bar\nu_\ell\nu_\tau)\bar\nu_\tau$},''
\href{http://arxiv.org/abs/1602.07671}{{\ttfamily arXiv:1602.07671 [hep-ph]}}.

\bibitem{Alok:2016qyh}
A.~K. Alok, D.~Kumar, S.~Kumbhakar, and S.~U. Sankar, ``{$D^*$ polarization as
  a probe to discriminate new physics in $B \to D^* \tau \bar \nu$},''
\href{http://arxiv.org/abs/1606.03164}{{\ttfamily arXiv:1606.03164 [hep-ph]}}.

\bibitem{Bardhan:2016uhr}
D.~Bardhan, P.~Byakti, and D.~Ghosh, ``{A closer look at the $R_D$ and
  $R_{D^*}$ anomalies},''
\href{http://arxiv.org/abs/1610.03038}{{\ttfamily arXiv:1610.03038 [hep-ph]}}.

\bibitem{Fan:2015kna}
Y.-Y. Fan, Z.-J. Xiao, R.-M. Wang, and B.-Z. Li, ``{The $B\to D^{(*)} l\nu_l$
  decays in the pQCD approach with the Lattice QCD input},''
\href{http://arxiv.org/abs/1505.07169}{{\ttfamily arXiv:1505.07169 [hep-ph]}}.

\bibitem{Becirevic:2013zw}
N.~Ko\v{s}nik, D.~Be\v{c}irevi\'c, and A.~Tayduganov, ``{Testing the Standard
  Model in $B \to D \tau \bar \nu$ decay with minimal theory input},''
  \href{http://arxiv.org/abs/1301.4037}{{\ttfamily arXiv:1301.4037 [hep-ph]}}.
[PoSConfinementX,244(2012)].

\bibitem{pdg2014}
{\bfseries Particle Data Group} Collaboration, K.~A. Olive {\em et~al.},
  ``{Review of Particle Physics},''
\href{http://dx.doi.org/10.1088/1674-1137/38/9/090001}{{\em Chin. Phys.}
  {\bfseries C38} (2014) 090001}.

\bibitem{Bordone:2016tex}
M.~Bordone, G.~Isidori, and D.~van Dyk, ``{Impact of leptonic $\tau $ decays on
  the distribution of $B\rightarrow P\mu \bar{\nu }$ decays},''
  \href{http://dx.doi.org/10.1140/epjc/s10052-016-4202-x}{{\em Eur. Phys. J.}
  {\bfseries C76} no.~7, (2016) 360},
\href{http://arxiv.org/abs/1602.06143}{{\ttfamily arXiv:1602.06143 [hep-ph]}}.

\bibitem{Ligeti:2016npd}
Z.~Ligeti, M.~Papucci, and D.~J. Robinson, ``{New Physics in the Visible Final
  States of $B\to D^{(*)}\tau\nu$},''
\href{http://arxiv.org/abs/1610.02045}{{\ttfamily arXiv:1610.02045 [hep-ph]}}.

\bibitem{Chang:2016cdi}
Q.~Chang, J.~Zhu, X.-L. Wang, J.-F. Sun, and Y.-L. Yang, ``{Study of
  semileptonic B¯⁎→Pℓν¯ℓ decays},''
  \href{http://dx.doi.org/10.1016/j.nuclphysb.2016.06.016}{{\em Nucl. Phys.}
  {\bfseries B909} (2016) 921--933},
\href{http://arxiv.org/abs/1606.09071}{{\ttfamily arXiv:1606.09071 [hep-ph]}}.

\bibitem{Bernardoni:2014kla}
{\bfseries ALPHA} Collaboration, F.~Bernardoni, J.~Bulava, M.~Donnellan, and
  R.~Sommer, ``{Precision lattice QCD computation of the $B^*B\pi$ coupling},''
  \href{http://dx.doi.org/10.1016/j.physletb.2014.11.051}{{\em Phys. Lett.}
  {\bfseries B740} (2015) 278--284},
\href{http://arxiv.org/abs/1404.6951}{{\ttfamily arXiv:1404.6951 [hep-lat]}}.

\bibitem{Flynn:2015xna}
{\bfseries RBC, UKQCD} Collaboration, J.~M. Flynn, P.~Fritzsch, T.~Kawanai,
  C.~Lehner, B.~Samways, C.~T. Sachrajda, R.~S. Van~de Water, and O.~Witzel,
  ``{The $B^*B\pi$ Coupling Using Relativistic Heavy Quarks},''
  \href{http://dx.doi.org/10.1103/PhysRevD.93.014510}{{\em Phys. Rev.}
  {\bfseries D93} no.~1, (2016) 014510},
\href{http://arxiv.org/abs/1506.06413}{{\ttfamily arXiv:1506.06413 [hep-lat]}}.

\bibitem{ElBennich:2010ha}
B.~El-Bennich, M.~A. Ivanov, and C.~D. Roberts, ``{Strong $D^* \to D \pi$ and
  $B^* \to B \pi$ couplings},''
  \href{http://dx.doi.org/10.1103/PhysRevC.83.025205}{{\em Phys. Rev.}
  {\bfseries C83} (2011) 025205},
\href{http://arxiv.org/abs/1012.5034}{{\ttfamily arXiv:1012.5034 [nucl-th]}}.

\bibitem{FloresTlalpa:2005fz}
A.~Flores-Tlalpa, G.~L\'opez~Castro, and G.~Sanchez~Toledo, ``{Radiative
  two-pion decay of the tau lepton},''
  \href{http://dx.doi.org/10.1103/PhysRevD.72.113003}{{\em Phys. Rev.}
  {\bfseries D72} (2005) 113003},
\href{http://arxiv.org/abs/hep-ph/0511315}{{\ttfamily arXiv:hep-ph/0511315
  [hep-ph]}}.

\bibitem{Dungel:2010uk}
{\bfseries Belle} Collaboration, W.~Dungel {\em et~al.}, ``{Measurement of the
  form factors of the decay $B^0 \to D^{*-} \ell^+ \nu_\ell$ and determination
  of the CKM matrix element $|V_{cb}|$},''
  \href{http://dx.doi.org/10.1103/PhysRevD.82.112007}{{\em Phys. Rev.}
  {\bfseries D82} (2010) 112007},
\href{http://arxiv.org/abs/1010.5620}{{\ttfamily arXiv:1010.5620 [hep-ex]}}.

\end{thebibliography}\endgroup

\end{document}